\documentclass[referee]{cjaa}           

\usepackage{graphicx}                   
\input{epsf.sty}                        
\input{psfig.sty}                       

\newcommand{\mo}    {M_{\odot}}

\newcommand{\pri}   {${\rlap.}^{\prime \prime}$}
\newcommand{\rl}    {${\rlap.}^{\rm s}$}

\begin{document}

   \title{Microquasars as high-energy gamma-ray sources
}

   \volnopage{Vol.0 (200x) No.0, 000--000}      
   \setcounter{page}{1}          

   \author{Josep M. Paredes
      \inst{}\mailto{}
      }
   \offprints{A.-Y. Zhou}                   

   \institute{Departament d'Astronomia i Meteorologia, Universitat de
Barcelona, Av. Diagonal 647, 08028 Barcelona, Spain \\
             \email{jmparedes@ub.edu}
          }

   \date{Received~~2004 month day; accepted~~2004~~month day}

   \abstract{Galactic microquasars are certainly one of the most recent
additions to the field of high energy astrophysics and have attracted
increasing interest over the last decade. However, the high energy part of the
spectrum of microquasars is the most poorly known, mainly due the lack of
sensitive instrumentation in the past. Microquasars are now primary targets
for all of the observatories working in the X-ray and gamma-ray domains. They 
also appear as the possible counterparts for some of the unidentified sources
of high-energy gamma-rays detected by the experiment EGRET on board the
satellite COMPTON-GRO. This paper provides a general review of the main
observational results obtained up to now as well as a summary of the scenarios
for production of high-energy $\gamma$-rays at the present moment. 
 } 

   \authorrunning{J.M. Paredes}            
   \titlerunning{Microquasars as high-energy gamma-ray sources}  

   \maketitle

%
%
\section{Introduction}           

\label{sect:intro}

The spectrum of microquasars extends from radio wavelengths to $\gamma$-rays
(Mirabel \& Rodr\'{\i}guez \cite{mir99}, Fender \cite{fen04}). Therefore, a
complete understanding of these objects necessarily requires a
multi-wavelength approach using both ground and space-based telescopes and
detectors. Such an effort is justified in view of the high interest raised by
the quasar-microquasars connections. Microquasars provide an excellent
laboratory for a suitable study of mass accretion and ejection phenomena in
the strong gravitational field of a black hole or neutron star on human
timescales. For instance, it has been possible to prove the direct connection
between instabilities of the accretion disk and formation of jets in the
microquasar and black hole candidate GRS~1915+105 on timescales of
minutes-hours (Fender et~al. \cite{fen97}, Mirabel et~al. \cite{mir98}). On
the other hand, microquasars appear as a possible explanation for some of the
unidentified sources of high-energy $\gamma$-rays detected by the experiment
EGRET on board the satellite COMPTON-GRO. In particular, the microquasar
LS~5039 is the best representative of the proposed connection between
microquasars and unidentified EGRET sources (Paredes et~al.~\cite{par00}),
although it is not a unique case. LS I+61 303, whose microquasar nature was
stablished when VLBI observations detected a relativistic radio jet (Massi et
al. \cite{mas01}), is also the possible counterpart of a high-energy
$\gamma$-ray source (Gregory \& Taylor \cite{greta78}, Kniffen
et~al.~\cite{kni97}). So, microquasars have turned out to be likely
high-energy $\gamma$-ray emitting objects, being also interesting targets to
be observed by the new generation of Cherenkov imaging telescopes at very
high-energy $\gamma$-rays.


\section{The microquasar population in our Galaxy}
\label{sect:popul}

Microquasars, understood as X-ray Binaries with relativistic radio jets,
represent a growing subset of the X-ray binary population in the Galaxy. The
most recent catalogue of High Mass X-ray Binaries (HMXBs) contains 130 sources
(Liu et~al.~\cite{liu00}), while the catalogue of Low Mass X-ray Binaries
(LMXBs) amounts to 150 objects (Liu et~al.~\cite{liu01}). Considering both
catalogues together, there are about 75 X-ray pulsars, which are not radio
emitters, and a total of 43 radio emitting sources, some of which have been
found to be microquasars (Rib\'o \cite{rib02}, Rib\'o \cite{rib04}). Recently,
it has been estimated that the total number of X-ray Binaries in the Galaxy
brighter than 2$\times 10^{34}$ erg~s$^{-1}$ is about 705, being distributed
as $\sim$325 LMXBs and $\sim$380 HMXBs (Grimm et~al.~\cite{gri02}). This
suggests an upper limit on the population of microquasars in the Galaxy of
about one hundred systems. A decade ago, one could count the known
microquasars with the fingers of one hand. Now, the situation is much better
but we do not have yet a microquasar population big enough from which
statistically robust results can be derived.

At the time of writing, a total of 15 microquasar systems have been
identified. All of them are listed in Table~\ref{census}. The top part of the
table is reserved for High Mass X-ray Binaries, while the bottom part contains
those of low mass. Within each group, the objects are sorted by right
ascension and the following information is given: name and position; type of
system; distance; orbital period; mass of the compact object; degree of
activity (persistent/transient radio emission); apparent velocity of the
ejecta; inclination and size of the jets; remarks.

There are some sources, e.g. XTE~J1118+480, considered as microquasar
candidates because their radio emission has not yet been resolved in
relativistic jets, although their existence is inferred from a theoretical
point of view.

It is very likely that discoveries in the near future will increase
substantially the galactic microquasar census and Table~\ref{census} will soon
become obsolete. Similarly, nothing prevents us from thinking that
microquasars in nearby galaxies could eventually be identified. However, such
a detection will certainly be quite difficult due to the very large distances
involved. Of course, such detection could become feasible provided that
extreme relativistic beaming effects can occur (e.g. microblazars as possible
ultraluminous X-ray sources).

\begin{table}
{\small
\caption[]{\label{census} {\bf Microquasars in our Galaxy}}
\begin{tabular}{@{}l@{\hspace{0.07cm}}@{}l@{\hspace{0.07cm}}c@{\hspace{0.05cm}}cc@{\hspace{0.05cm}}c@{\hspace{0.05cm}}c@{\hspace{0.05cm}}c@{\hspace{0.05cm}}c@{\hspace{0.05cm}}c@{}c@{}}
             &                 &           &      &              &              &                           \\ 
\hline \noalign{\smallskip}
Name  & Position &  System  & $D$ & $P_{\rm orb}$  & $M_{\rm compact}$   & Activity &
$\beta_{\rm apar}$ & $\theta$$^{\rm (c)}$ & Jet size & Remarks$^{\rm (d)}$ \\
& (J2000.0)  & type$^{\rm (a)}$& (kpc) & (d)   &   $(\mo)$  & radio$^{\rm (b)}$&   &  & (AU)\\ 
\noalign{\smallskip} \hline \noalign{\smallskip}
\multicolumn{10}{c}{\bf High Mass X-ray Binaries (HMXB)}\\
\noalign{\smallskip} \hline \noalign{\smallskip}

{\bf LS~I~+61~303} & $02^{\rm h}40^{\rm m}$31\rl 66 &B0V & 2.0 & 26.5 & $-$ & p & $\geq$0.4 & $-$ & 10$-$700 & Prec?\\
& $+61^{\circ}13^{\prime}$45\pri 6 & +NS? &  & & & & & \\

{\bf V4641~Sgr} & $18^{\rm h}19^{\rm m}$21\rl 48  & B9III & $\sim10$ & 2.8  & 9.6 & t & $\ge9.5$ & $-$&$-$ & \\
& $-25^{\circ}25^{\prime}$36\pri 0 &+BH & \\
 
{\bf LS~5039} & $18^{\rm h}26^{\rm m}$15\rl 05   &  O6.5V((f)) & 2.9 & 4.4 & 1$-$3 & p  & $\geq0.15$ &$<81^{\circ}$& 10$-$1000 & Prec?\\
&$-14^{\circ}50^{\prime}$54\pri 24 & +NS?& &   & & & & &  \\
  
{\bf SS~433} & $19^{\rm h}11^{\rm m}$49\rl 6&evolved A?   & 4.8  &  13.1 &  11$\pm$5?& p & 0.26 &  $79^{\circ}$&$\sim10^4$$-$$10^6$ & Prec   \\
& $+04^{\circ}58^{\prime}58^{\prime\prime}$ &+BH?  & &   & & & & & &XRJ\\

{\bf Cygnus~X-1} & $19^{\rm h}58^{\rm m}$21\rl 68&O9.7Iab  & 2.5 & 5.6 & 10.1 &  p &$-$& 40$^{\circ}$& $\sim40$\\
& $+35^{\circ}12^{\prime}$05\pri 8& +BH & &   & & & & & \\
 
{\bf Cygnus~X-3} & $20^{\rm h}32^{\rm m}$25\rl 78 &WNe     &  9      &  0.2   & $-$& p  & 0.69 & 73$^{\circ}$ & $\sim10^4$    \\
&$+40^{\circ}57^{\prime}$28\pri 0 & +BH? & &   & & & & & \\

\noalign{\smallskip} \hline \noalign{\smallskip}
\multicolumn{10}{c}{\bf Low Mass X-ray Binaries (LMXB)}\\
\noalign{\smallskip} \hline \noalign{\smallskip}

      
{\bf Circinus~X-1}    & $15^{\rm h}20^{\rm m}$40\rl 9& Subgiant &  5.5     &  16.6  &$-$&  t   & $>15$& $<6^{\circ}$ & $>10^4$ \\
& $-57^{\circ}10^{\prime}01^{\prime\prime}$ & +NS \\
 
{\bf XTE~J1550$-$564} & $15^{\rm h}50^{\rm m}$58\rl 70 & G8$-$K5V & 5.3  & 1.5   & 9.4 &  t
&$>2$& $-$& $\sim10^3$  & XRJ   \\
& $-56^{\circ}28^{\prime}$35\pri 2& +BH & &   & & & & &  \\
 
{\bf Scorpius~X-1}  &  $16^{\rm h}19^{\rm m}$55\rl 1 & Subgiant    & 2.8      &  0.8  &  1.4 &  p &$ 0.68$& $44^{\circ}$& $\sim40$    \\
 & $-15^{\circ}38^{\prime}25^{\prime\prime}$& +NS  \\
  
{\bf GRO~J1655$-$40} & $16^{\rm h}54^{\rm m}$00\rl 25 & F5IV  & 3.2    &  2.6   & 7.02 & t & 1.1& $72^{\circ}$$-$$85^{\circ}$&  8000  & Prec?   \\
& $-39^{\circ}50^{\prime}$45\pri 0 & +BH & &   & & & & &  \\
   
{\bf GX~339$-$4}   & $17^{\rm h}02^{\rm m}$49\rl 5 & $-$         & $>6$     &  1.76  & 5.8$\pm$0.5 & t& $-$&$-$& $<$ 4000    \\
& $-48^{\circ}47^{\prime}23^{\prime\prime}$& +BH \\

{\bf 1E~1740.7$-$2942}& $17^{\rm h} 43^{\rm m} 54$\rl 83&$-$  & 8.5? &  12.5?   &$-$ & p &$-$& $-$&$\sim10^6$  \\
& $-29^{\circ} 44^{\prime}$42\pri 60& +BH ?& &   & & & & & \\

{\bf XTE~J1748$-$288} & $17^{\rm h}48^{\rm m}$05\rl 06& $-$ &  $\geq8$   &  ?    &   $>4.5$? & t & 1.3&$-$ & $>10^4$         \\
& $-28^{\circ}28^{\prime}$25\pri 8 &+BH? \\
    
{\bf GRS~1758$-$258}  &$18^{\rm h}01^{\rm m}$12\rl 40  & $-$ & 8.5?  & 18.5?  &$-$   & p &$-$& $-$&$\sim10^6$\\
& $-25^{\circ}44^{\prime}$36\pri 1 &+BH ?\\
   
{\bf GRS~1915+105} &  $19^{\rm h}15^{\rm m}$11\rl 55 &  K$-$M III   & 12.5 & 33.5  & 14$\pm$4 &  t & 1.2$-$1.7& $66^{\circ}$$-$$70^{\circ}$&$\sim10$$-$$10^4$ & Prec?\\
&$+10^{\circ}56^{\prime}$44\pri 7 &+BH & &   & & & & & \\
\hline
\end{tabular}
{\small
Notes: $^{\rm (a)}$ NS: neutron star; BH: black hole. $^{\rm (b)}$ p: persistent; t:transient. $^{\rm (c)}$ jet inclination.
$^{\rm (d)}$ Prec: precession; XRJ: X-ray jet.}}
\end{table}

\section{Why should we expect microquasars to be $\gamma$-ray emitters?}
\label{sect:gamma}

It is widely accepted that relativistic jets in AGNs are strong emitters of
$\gamma$-rays with GeV energies (e.g. von Montigny et~al.~\cite{mon95}).
Generally speaking, and allowing for their similarity (Mirabel \&
Rodr\'{\i}guez~\cite{mir99}), one could also expect the jets in microquasars
to be GeV $\gamma$-ray emitters. In some cases, however, the sensitivity of
the current $\gamma$-ray detectors may not be high enough to detect such
emission. For instance, based on the physical parameters derived from
observations of outbursts, the expected $\gamma$-ray flux of GRS~1915+105 up
to very high-energy $\gamma$-rays has been estimated from inverse Compton
scattering of the synchrotron photons (Atoyan \& Aharonian~\cite{ato99}). The
resulting fluxes could have been hardly detected by EGRET, being also of
transient nature, but they are within the sensitivity of the future AGILE and
GLAST missions, about 10--100 times better than that of EGRET. 


Several models have been developed to explore the high energy emission from
the jets of microquasars. Two kinds of model can be found in the literature
depending on whether hadronic or leptonic jet matter dominates the emission at
such an energy range: the hadronic jet models (e.g. Romero et~al.
\cite{rom03}), and the leptonic jet models. Among leptonic jet models, there
are IC jet emission models that can produce X-rays and $\gamma$-rays, based in
some cases on the synchrotron self-Compton (SSC) process (i.e. Band \&
Grindlay \cite{Band&grindlay86}; Atoyan \& Aharonian \cite{ato99}), and in
other cases on external sources for the IC seed photons (EC) (i.e. Kaufman
Bernad\'o et~al. \cite{Kaufman02}; Georganopoulos et~al.
\cite{Georganopoulos02}). In addition, there are synchrotron jet emission
models that can produce X-rays (i.e. Markoff et~al. \cite{mar03}). A general
description of such models can be found in Romero~(\cite{rom04b}).

\section{High energy observations}

In the next subsections we review and comment on the observational data of the
microquasars listed in Table~\ref{census} at energies from soft to very
high-energy $\gamma$-rays, and we quote them in Table~\ref{detections}.

\begin{table}[t!]
{\small
\caption[]{\label{detections} {\bf High energy emission from microquasars}}
\begin{tabular}{@{}l@{\hspace{0.07cm}}c@{\hspace{0.05cm}}cc@{\hspace{0.05cm}}c@{\hspace{0.05cm}}c@{\hspace{0.05cm}}c@{\hspace{0.05cm}}c@{\hspace{0.05cm}}c@{}}
             &                 &           &      &              &              &                           \\ 
\hline \noalign{\smallskip}
Name  &  \multicolumn{2}{c} {INTEGRAL$^{\rm (a)}$}   & \multicolumn{2}{c} {BATSE$^{\rm (b)}$} 
&     COMPTEL$^{\rm (c)}$ & EGRET$^{\rm (d)}$ & Others$^{\rm (e)}$  \\
& 30$-$50 keV & 40$-$100 keV & 20$-$100 keV   &   160$-$430 keV & 1$-$30 MeV &  $>$ 100 MeV &  & \\ 
& (significance) & (count/s) & (significance) & (mCrab) & (GRO) & (3EG)\\
\noalign{\smallskip} \hline \noalign{\smallskip}
\multicolumn{9}{c}{\bf High Mass X-ray Binaries (HMXB)}\\
\noalign{\smallskip} \hline \noalign{\smallskip}

{\bf LS~I~+61~303} & $-$ & $-$ & 5.2 & 5.1$\pm$2.1  & J0241+6119? & J0241+6103? &  \\

{\bf V4641~Sgr}  & $-$ & $-$ & $-$  & $-$ & $-$ & $-$ &  \\

{\bf LS~5039} &  $-$ & $-$ & 10.7 & 3.7$\pm$1.8 & J1823$-$12?  & J1824$-$1514? & \\
  
{\bf SS~433} & 13.5  & $<$1.02  &  21.7 &  0.0$\pm$2.8 & $-$ & $-$ &   \\
  
{\bf Cygnus~X-1} & 676.6  & 66.4$\pm$0.1 & 1186.8 & 924.5$\pm$2.5 &  yes &$-$& S\\
  
{\bf Cygnus~X-3} &  122.7     &  5.7$\pm$0.1      &  197.8   & 15.5$\pm$2.1&$-$   & $-$ & O, T?  \\

\noalign{\smallskip} \hline \noalign{\smallskip}
\multicolumn{9}{c}{\bf Low Mass X-ray Binaries (LMXB)}\\
\noalign{\smallskip} \hline \noalign{\smallskip}

      
{\bf Circinus~X-1}    &  $-$ &  $-$     &  3.8  & 0.3$\pm$2.6&  $-$   & $-$& \\
 
{\bf XTE~J1550$-$564} &   8.6 & 0.6$\pm$0.07  & 17.1   & $-$2.3$\pm$2.5 & $-$   & $-$&  \\
 
{\bf Scorpius~X-1}     & 111.6    & 2.3$\pm$0.1      &  460.6  &  9.9$\pm$2.2 &  $-$ & $-$ &   \\
  
{\bf GRO~J1655$-$40} & $-$  & $-$    &  40.6   & 23.4$\pm$3.9 & $-$ & $-$& O    \\
   
{\bf GX~339$-$4}   &   21.9       & 0.55$\pm$0.03    &  89.0  & 580$\pm$3.5 & $-$&$-$& S    \\ 
  
{\bf 1E~1740.7$-$2942}& 147.3  & 4.32$\pm$0.03 & 92.4  & 61.2$\pm$3.7 & $-$& $-$& S  \\
 
{\bf XTE~J1748$-$288} &  $-$ &  $-$  &   $-$12.4 & $-$ & $-$ & $-$  &S        \\

{\bf GRS~1758$-$258}  & 135.9 & 3.92$\pm$0.03  & 74.3  & 38.0$\pm$3.0  & $-$ &$-$& S\\
   
{\bf GRS~1915+105}    &  144.9  & 8.63$\pm$0.13 & 208.8  & 33.5$\pm$2.7 & $-$  & $-$ & S, T?\\

\hline
\end{tabular}
{\small
Notes: $^{\rm (a)}$ The first IBIS/ISGRI soft gamma-ray galactic plane survey catalog 
(Bird et al. 2004). $^{\rm (b)}$~BATSE Earth occultation catalog, Deep sample results 
(Harmon et al. 2004). 
$^{\rm (c)}$ The first COMPTEL source catalogue (Sch\"onfelder et~al.~\cite{sch00})
$^{\rm (d)}$ The third EGRET catalog of high-energy $\gamma$-ray sources 
(Hartman et al.~\cite{har99})
$^{\rm (e)}$ S: SIGMA instrument onboard GRANAT satellite; O: OSSE;
T: TeV source
}
}
\end{table}

\subsection{INTEGRAL, BATSE and COMPTEL sources}

Bird et al.~(\cite{bir04}) have reported the first high-energy survey catalog
obtained with the IBIS $\gamma$-ray imager on board INTEGRAL, covering the
first year data. This initial survey has revealed the presence of $\sim$120
sources detected with a good sensitivity in the energy range 20$-$100~keV.
Among the detected sources we have inspected the microquasars listed in
Table~\ref{census}. In the second column of Table~\ref{detections} we list
their significance in the 30$-$50~keV energy range and in the third column
their flux (count/s) and error or upper limit in the energy range of
40$-$100~keV.

The Burst and Transient Source Experiment (BATSE), aboard the Compton Gamma
Ray Observatory (CGRO), monitored the high energy sky using the Earth
occultation technique (EOT). A compilation of BATSE EOT observations has been
published recently (Harmon et al.~\cite{har04}), with the flux data for the
sample being presented in four energy bands. From this catalog we have
selected also the data on microquasars. In the fourth column of
Table~\ref{detections}, the significance in the energy range 20$-$100~keV is
listed, while in the fifth column we have listed their flux in the energy
range 160$-$430~keV in mCrab units. Cygnus~X-1 and Cygnus~X-3 have been
studied extensively by BATSE.

The instrument COMPTEL, also aboard the CGRO, detected 32 steady sources and
31 $\gamma$-ray bursters (Sch\"onfelder et~al.~\cite{sch00}). Among the
continuum sources detected there are the microquasar Cygnus~X-1 and other two
sources, GRO~J1823$-$12 and GRO~J0241+6119, possibly associated with two other
microquasars. See sixth column in Table~\ref{detections}. 

The standard interpretation of the emission in the low-energy $\gamma$-ray
range is that disc blackbody photons are Comptonized by thermal/nonthermal
electrons. There are state transitions (hard and low states) thought to be
related to changes in the mass accretion rate. Nevertheless, it is still
unclear whether this is what really happens. Alternatively, some groups have
suggested that this emission could come from the jet, basing this idea on
recent observational and theoretical results (see, i.e., Fender et~al.
\cite{fen03}, Markoff et~al. \cite{mar03}, Georganopoulos et~al.
\cite{Georganopoulos02}).

\subsection{EGRET sources}

 \begin{figure}[t!]
  \begin{minipage}[t]{0.5\linewidth}
  \centering
  \includegraphics[width=\textwidth]{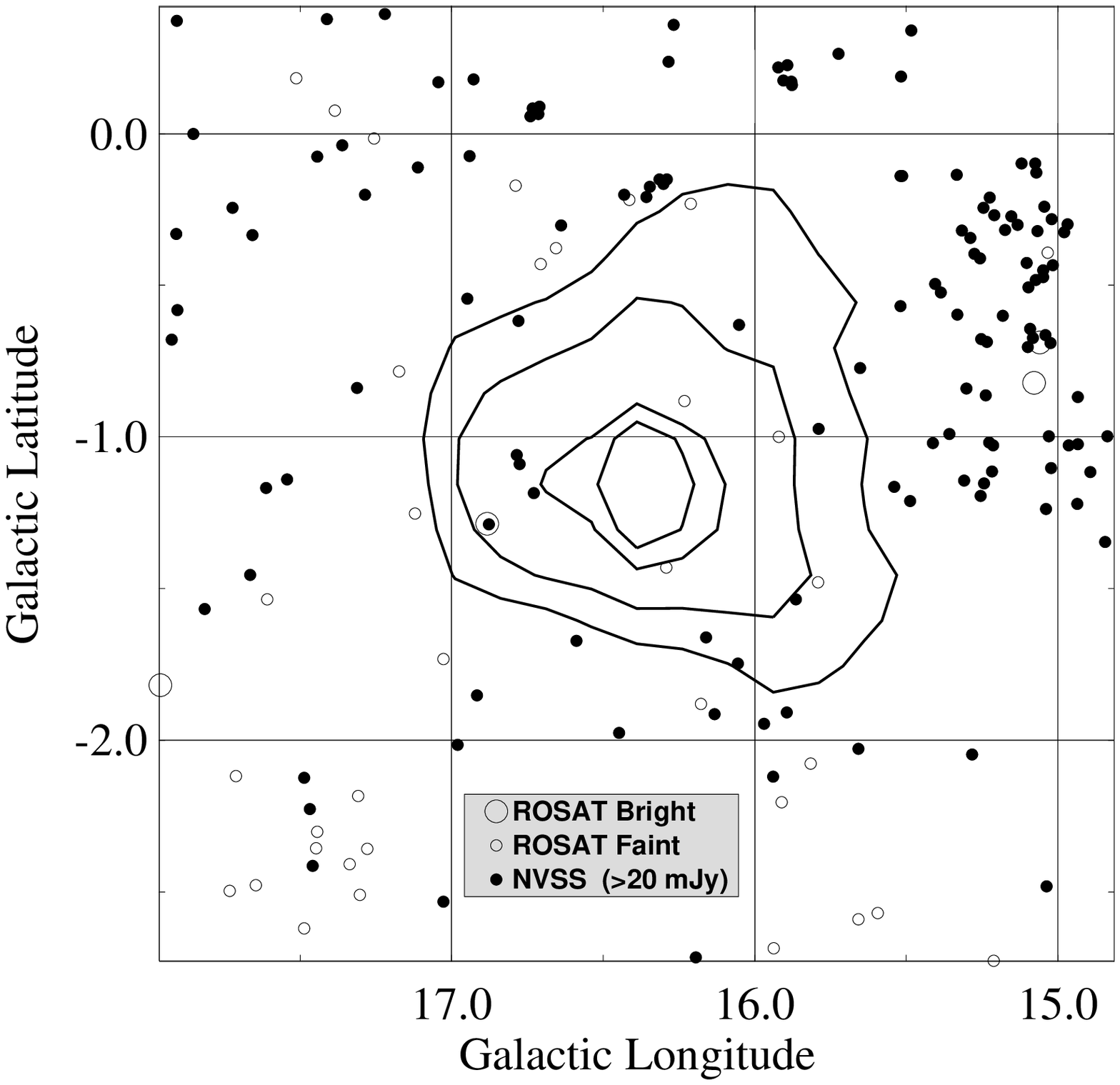}
  \vspace{-5mm}
  \caption{{\small Location map of the EGRET source 3EG~J1824$-$1514. The contours represent
  50\%, 68\%, 95\% and 99\% statistical probability that a $\gamma$-ray source
  lies within the given contour. The only X-ray and radio source  
  is LS~5039 (filled circle inside an open big circle at l=16.66$^{\circ}$ and 
  b=$-1.29^{\circ}$) (Rib\'o~\cite{rib02}).}}
  \label{Fig:EG1824}
  \end{minipage}
  \begin{minipage}[t]{0.5\textwidth}
  \centering
  \includegraphics[width=\textwidth]{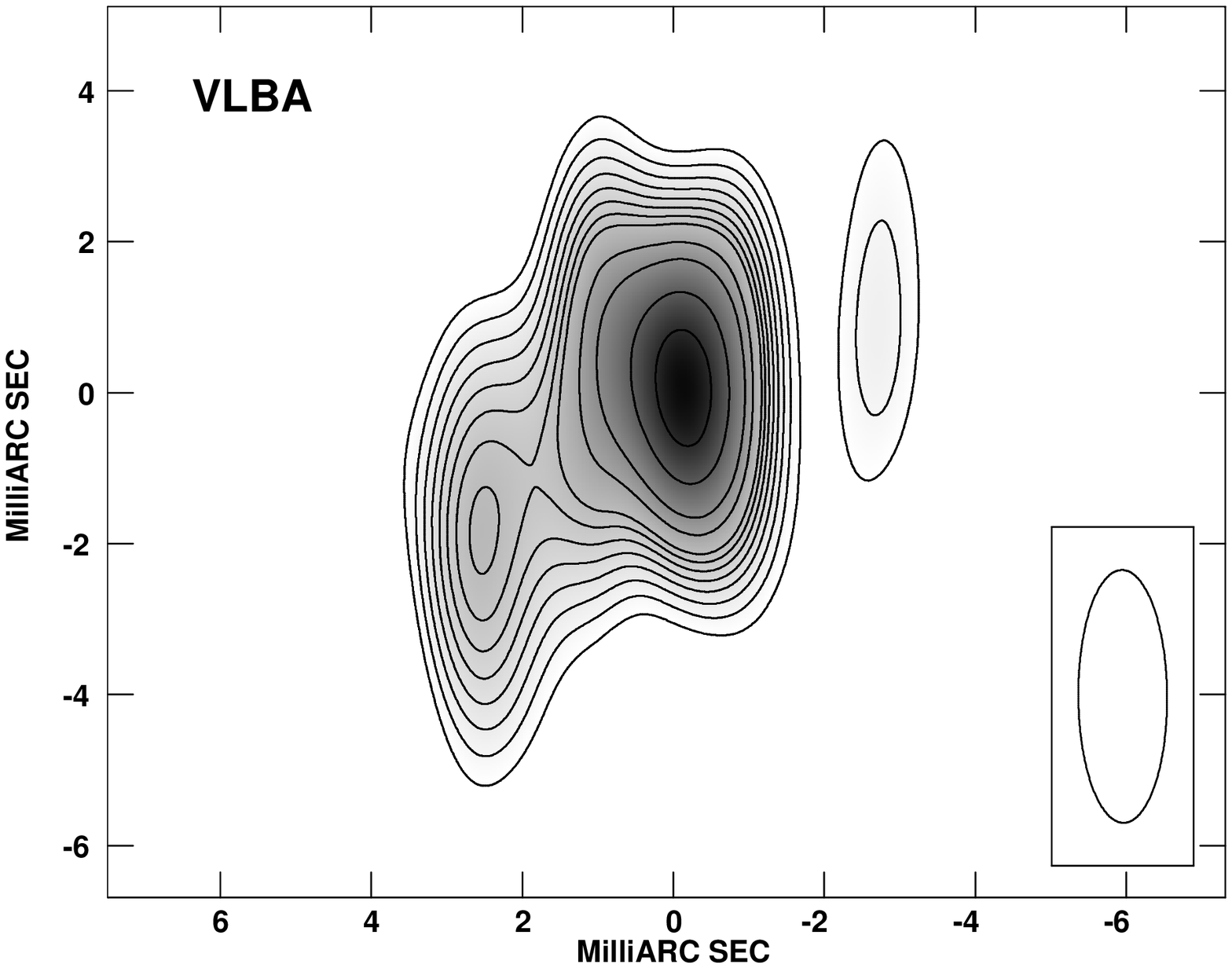}
  \vspace{-5mm}
  \caption{{\small Relativistic radio jets of LS~5039, observed with the VLBA, which reveal
  its microquasar nature (Paredes et~al.~\cite{par00}).}}
  \label{Fig:vlbajet}
  \end{minipage}
\end{figure}

\subsubsection{LS~5039 / 3EG~J1824$-$1514}

\begin{figure}[t!]
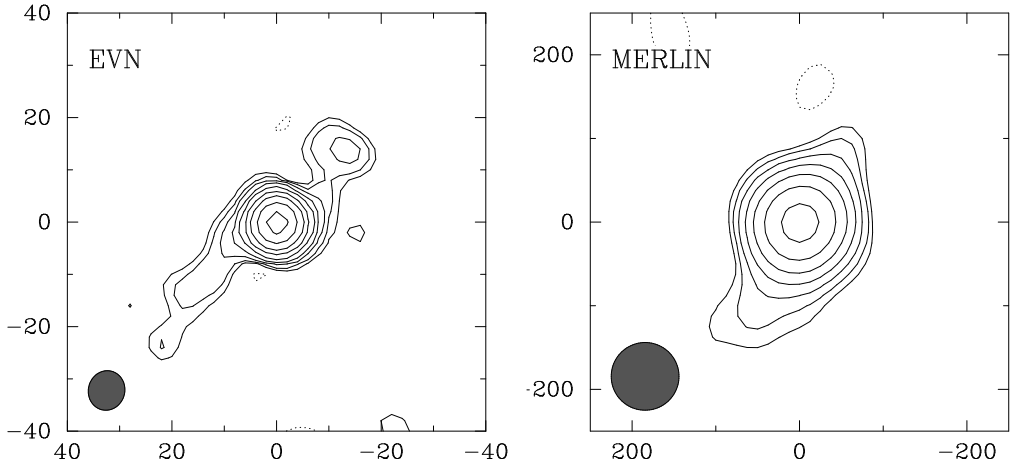

  \begin{minipage}[t]{0.5\linewidth}
  \centering
  \includegraphics[width=\textwidth]{ls5039_evn.eps}
  \vspace{-5mm}
  \end{minipage}%
  \begin{minipage}[t]{0.5\textwidth}
  \centering
  \includegraphics[width=\textwidth]{ls5039_merlin.eps}
  \end{minipage}%
  \caption{{\small Self-calibrated images of LS~5039 at 5~GHz obtained on March 1 2000
  with the EVN (left) and MERLIN (right). Axis units are in milliarc-seconds (Paredes et 
  al.~\cite{par02}).}}
  \label{Fig:jets}
\end{figure}

The discovery of the microquasar LS~5039, and its possible association with a
high-energy $\gamma$-ray source ($E>$100~MeV), provides observational evidence
that microquasars could also be sources of high-energy $\gamma$-rays (Paredes
et~al.~\cite{par00}). It is important to point out that this is the first time
that an association between a microquasar and a high-energy $\gamma$-ray
source has been reported. This finding opens up the possibility that other
unidentified EGRET sources could also be microquasars. LS~5039 is the only
X-ray source from the bright ROSAT catalog whose position is consistent with
the high energy $\gamma$-ray source 3EG~J1824$-$1514. LS~5039 is also the only
object simultaneously detected in X-rays and radio (Fig.~\ref{Fig:EG1824}),
which displays bipolar radio jets at sub-arcsecond scales
(Fig.~\ref{Fig:vlbajet}). New observations conducted with the EVN and MERLIN
confirm the presence of an asymmetric two-sided jet reaching up to
$\sim$1000~AU on the longest jet arm (Fig.~\ref{Fig:jets}) (Paredes et
al.~\cite{par02}, Rib\'o~\cite{rib02}). 

\begin{figure}[ht]
  \centering
  {\includegraphics[width=6cm]{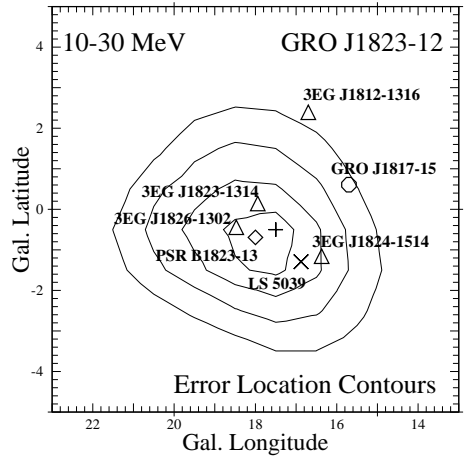}}
  \caption{Error location contours for the COMPTEL unidentified source
  GRO~J1823$-$12. The most likely source position (+) and possible 
  counterparts are overlaid (Collmar~\cite{col03}).}
  \label{Fig:comptel}
\end{figure}

Recently, Collmar~(\cite{col03}) has reported the detection of an unidentified
$\gamma$-ray source, GRO~J1823$-$12, at galactic coordinates
($l$=17.5$^{\circ}$, b=$-$0.5$^{\circ}$) by the COMPTEL experiment
(Fig.~\ref{Fig:comptel}). This source is among the strongest COMPTEL sources.
The source region, detected at high significance level, contains several
possible counterparts, being LS~5039 one of them. It is also worth noting that
BATSE has detected this source at soft $\gamma$-rays (see
Table~\ref{detections}). Taking into account these observational evidences,
from radio to high-energy $\gamma$-rays, LS~5039 appears to be a very likely
counterpart of the EGRET source 3EG~J1824$-$1514.

Regarding the theoretical point of view, the $\gamma$-ray emission from
3EG~J1824$-$1514, with a luminosity of L$_{\gamma} (>$100~MeV) $\sim$
10$^{35}$ erg~s$^{-1}$, is likely to be originated from inverse Compton effect
of the ultraviolet photons from a hot companion star scattered by the same
relativistic electrons responsible for the radio emission (Paredes
et~al.~\cite{par00}, Paredes et~al.~\cite{par02}). The energy shift in this
process is given by $E_\gamma\sim\gamma_{\rm e}^{2}E_{\rm ph}$, where the
energies of the $\gamma$-ray and the stellar photons are related by the
Lorentz factor of the electrons squared. For an O6.5 star in the main
sequence, such as the component of LS~5039, most of its luminosity is radiated
by photons with $E_{\rm ph}\sim$10~eV. To scatter them into $\gamma$-ray
photons with $E_\gamma\sim$100~MeV, electrons with a Lorentz factor of
$\gamma_{\rm e}\sim$10$^4$, or equivalently with energy $\sim$10$^{-2}$~erg,
are required. A detailed numerical model to explain the $\gamma$-ray emission
of LS~5039 has been proposed recently (Bosch-Ramon \& Paredes~\cite{bos04a}).

\subsubsection{LS~I~+61~303 / 3EG~J0241+6103}
\label{sect:lsi}

\begin{figure}[t!]
  \begin{minipage}[t]{0.5\linewidth}
  \centering
  \includegraphics[width=\textwidth]{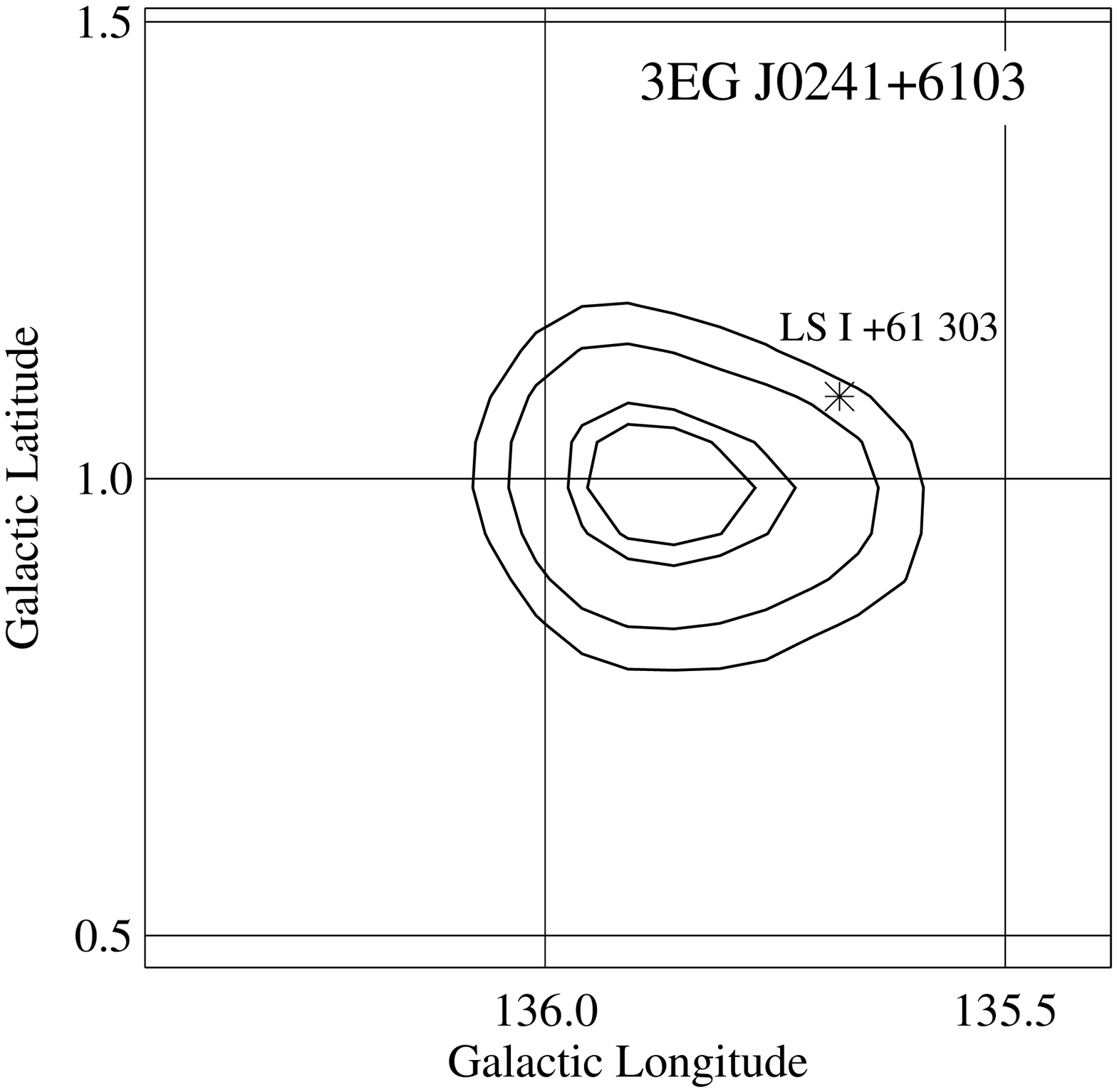}
  \vspace{-5mm}
  \caption{{\small Location map for 3EG~J0241+6103 in the 3rd EGRET catalog (adapted from Hartman et~al.~\cite{har99}).}}
  \label{Fig:EG0241}
  \end{minipage}%
  \begin{minipage}[t]{0.5\textwidth}
  \centering
  \includegraphics[width=\textwidth]{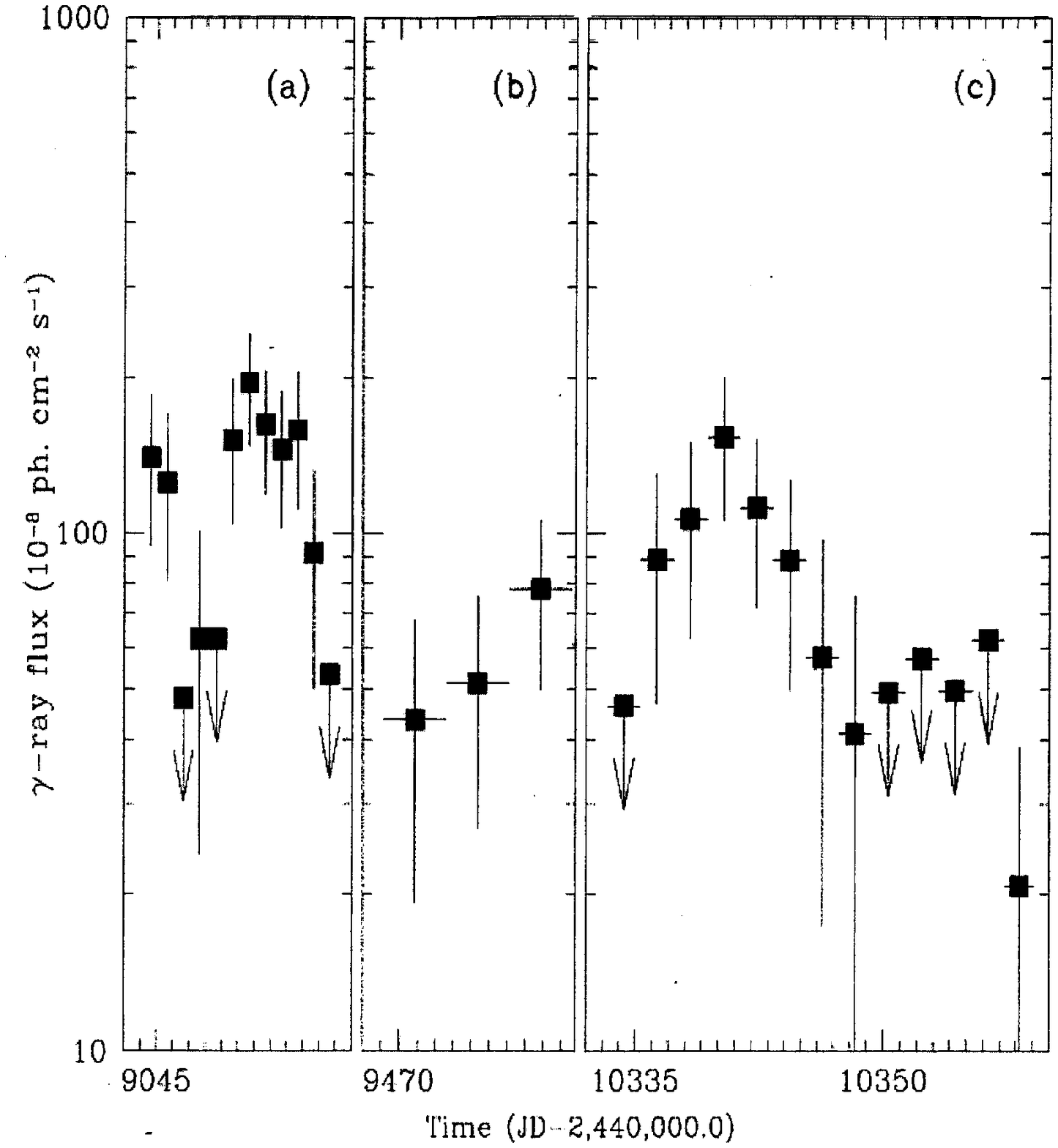}
  \vspace{-5mm}
  \caption{{\small EGRET light curve of 2CG~135+01 / 3EG~J0241+6103 for the best near-axis
  observations (Tavani et al.~\cite{tav98}).}}
  \label{Fig:gamvar}
  \end{minipage}%
\end{figure}

The well known High Mass X-ray binary LS~I~+61~303 has been classified as a
new microquasar after the discovery of relativistic jets (Massi
et~al.~\cite{mas01}, Massi et~al.~\cite{mas04}). This object has also been
proposed to be associated with the $\gamma$-ray source 2CG~135+01
(=3EG~J0241+6103) (Gregory \& Taylor \cite{greta78}, Kniffen
et~al.~\cite{kni97}). Although the broadband 1~keV--100~MeV spectrum of
LS~I~+61~303 remains uncertain, because OSSE and COMPTEL observations were
likely dominated by the quasar QSO~0241+622 emission, the EGRET angular
resolution is high enough to exclude this quasar as the source of the
$\gamma$-ray emission (Harrison et~al.~\cite{har00}).  BATSE marginally
detected the source, being the quasar also excluded as the origin of this
emission (see Table~\ref{detections}). We show the location map of
3EG~J0241+6103 in Fig.~\ref{Fig:EG0241}, where the position of LS~I~+61~303 is
indicated (Hartman et~al.~\cite{har99}). 

A timing analysis carried out recently by Massi~(\cite{mas04}) of pointed
EGRET observations (Tavani et al.~\cite{tav98}) (Fig.~\ref{Fig:gamvar})
suggests a period of 27.4$\pm$7.2 days, in agreement with the orbital period
of this binary system, of 26.496 days. This result, if confirmed, would
clearly support the association of LS~I~+61~303 with 3EG~J0241+6103.

\begin{figure}[ht!]
   \centering
   {\includegraphics[width=\textwidth]{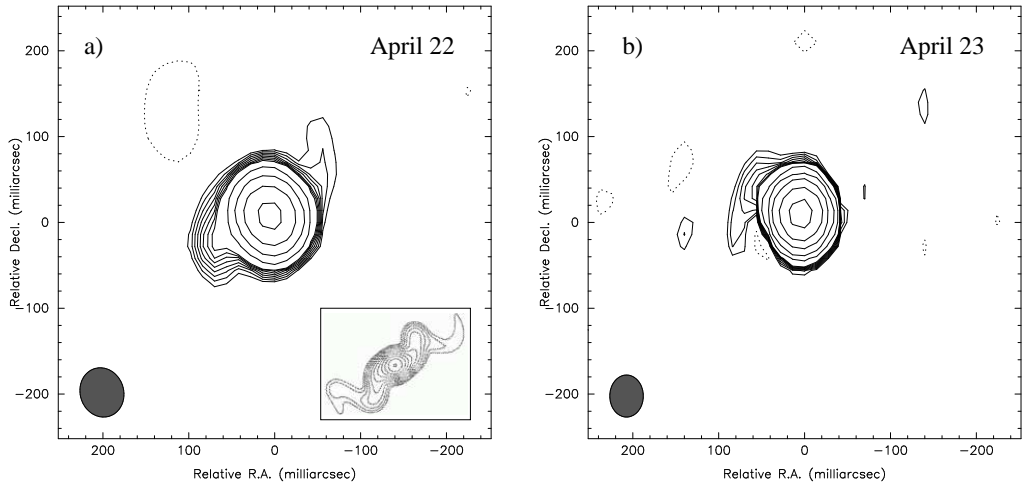}}   
   \caption{MERLIN self-calibrated images of LS~I~+61~303 at 5~GHz obtained on April 22
   (left) and April 23 (right). The S-shaped morphology recalls the precessing jet of
   SS~433, whose simulated radio emission is given in the small box (Massi et 
   al.~\cite{masal04}).}
   \label{Fig:prec}
   \end{figure}

This microquasar also seems to be a fast precessing system. MERLIN images
obtained in two consecutive days (Fig.~\ref{Fig:prec}) show a change in the
direction of the jets of about 50$^{\circ}$ that has been interpreted as a
fast precession of the system (Massi et al.~\cite{masal04}). If this is
confirmed, it could solve the puzzling VLBI structures observed so far, as
well as the short term variability of the associated $\gamma$-ray source
3EG~J0241+6103 (e.g. Wallace et~al.~\cite{wal00}). 

Up to now, the only existing radial velocity curve of LS~I~+61~303 was that
obtained by Hutchings and Crampton~(\cite{hut81}). Recently, after a
spectroscopic campaign, an improved estimation of the orbital parameters has
been obtained (Casares et~al.~\cite{cas04}). Here, we will just mention the
new high eccentricity (e=0.72$\pm$0.15) and the periastron orbital phase at
$\sim$0.2. These values are a key information for any interpretation of the
data obtained at any wavelength.

\begin{figure}[t!]
  \begin{minipage}[t]{0.5\linewidth}
  \centering
  \includegraphics[width=\textwidth]{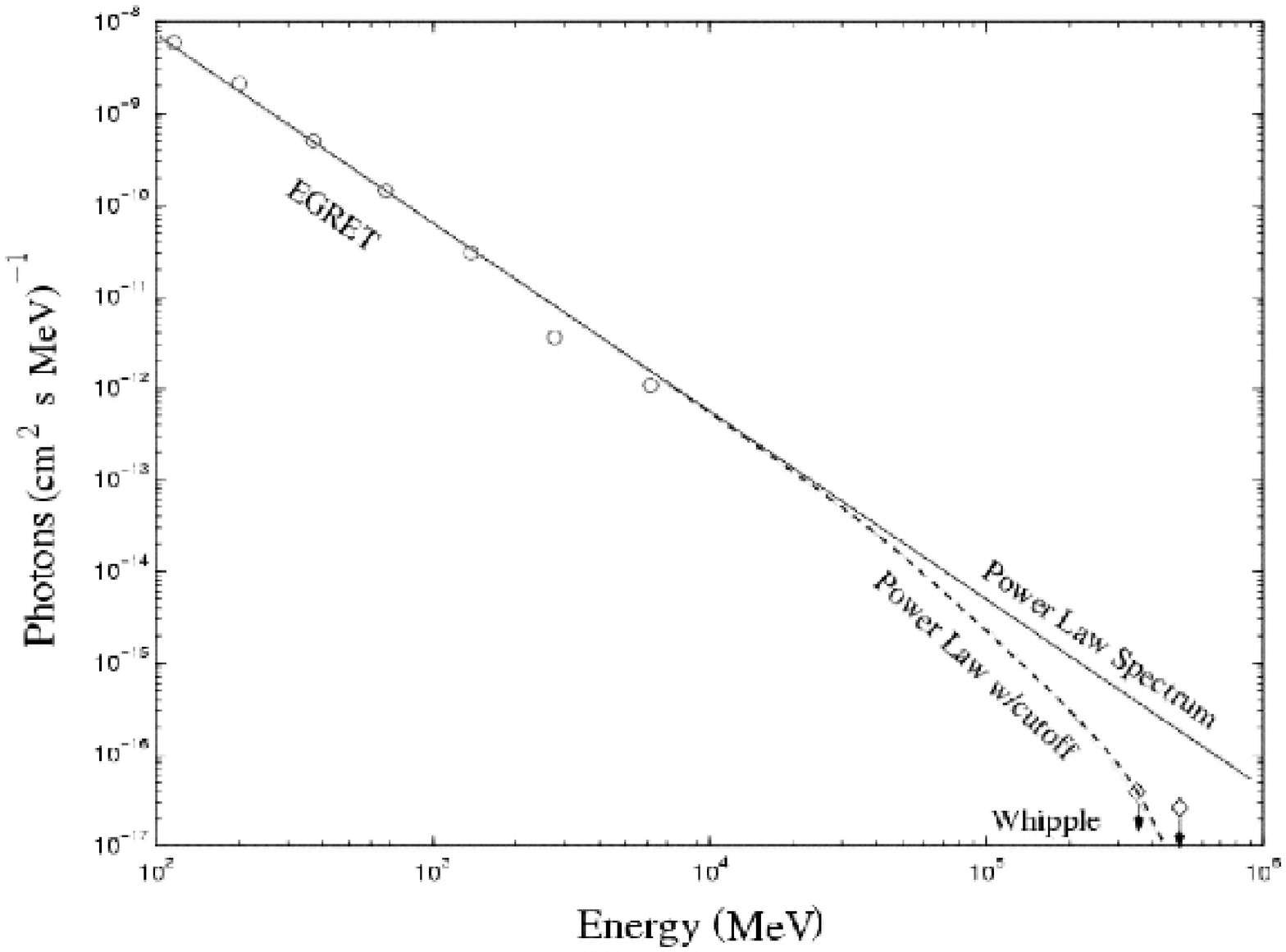}
  \vspace{-5mm}
  \caption{{\small EGRET data points of 3EG~J0241+6103 (circles) and the flux      upper limits
   (diamonds) obtained by Whipple (Hall et al.~\cite{hal03}).}}
  \label{Fig:whipple}
  \end{minipage}%
  \begin{minipage}[t]{0.5\textwidth}
  \centering
  \includegraphics[width=\textwidth]{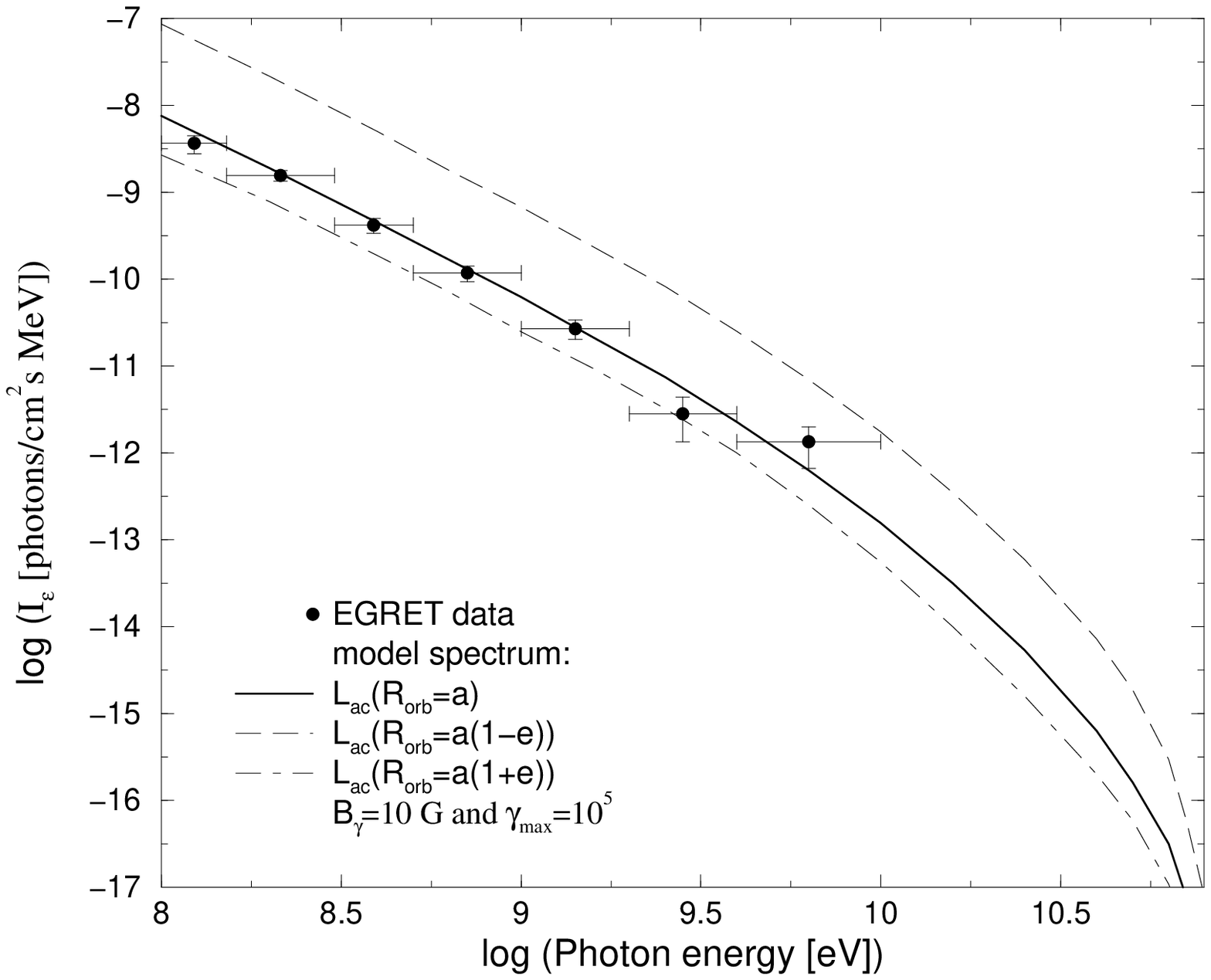}
  \vspace{-5mm}
  \caption{{\small Computed spectral photon distribution of LS~I~+61~303 above 100 MeV plotted with the
  EGRET data points of 3EG~J0241+6103 (Bosch-Ramon \& Paredes~\cite{bos04b}).}}
  \label{Fig:lsimodel}
  \end{minipage}%
\end{figure}

Hall et al.~(\cite{hal03}) gave upper limits for the emission associated to
LS~I~+61~303 / 3EG~J0241+6103 at very high-energy $\gamma$-rays from
observations performed by the Cherenkov telescope Whipple
(Fig~\ref{Fig:whipple}). Several models have been proposed to explore the high
energy emission of this source (e.g. Taylor et al.~\cite{tay96},
Punsly~\cite{pun99}, Harrison et al.~\cite{har00}, Leahy~\cite{lea04}). The
most recent model has been presented by Bosch-Ramon \&
Paredes~(\cite{bos04b}), who explore with a  detailed numerical model if this
system can produce the emission and to present the variability detected by
EGRET ($>$100~MeV). We reproduce in Fig.~\ref{Fig:lsimodel} the computed
spectral photon distribution, which is able to fit the data.

\subsubsection{Other EGRET/microquasar candidates}

After a multiwavelength study of the X-ray source AX~J1639.0$-$4642, Combi
et~al.~(\cite{com04}) have proposed that this source is a dust-enshrouded 
microquasar candidate. This source is within the 95\% location contours of
3EG~J1639$-$4702 and has been proposed by the same authors to be the 
counterpart of the EGRET source. If confirmed, this would be an important
result, although by now there are other possibilities such as the presence of
the pulsar PSR~J1637$-$4642 within the 95\% confidence contours (Torres 
et~al.~\cite{tor01}).

\subsubsection{EGRET candidate microquasars as runaway objects}

A hot recent topic in the microquasar field is the measurement of high spatial
velocities for some of these objects (the so-called runaway microquasars). 
Regarding the two microquasars that are EGRET source candidates, interesting
kinematic properties have been observed. LS~5039, with a systemic velocity of
150 km~s$^{-1}$, is escaping from its local environment with a very high
velocity component perpendicular to the Galactic Plane (Rib\'o et~al. 2002).
Such behaviour may be the result of the supernova explosion which created the
compact object in this binary system. According to the computed trajectory and
the possible lifetime of the donor, LS~5039 could reach a galactic latitude of
$-$12$^{\circ}$ and still behave as a microquasar. Taking into account the
possible association of this microquasar with the high-energy $\gamma$-ray
source 3EG~J1824$-$1514, one could expect to detect gamma-ray microquasars up
to $\pm$10 degrees of galactic latitudee. Also LS~I~+61~303 is running away
from its birth place with a linear momentum of 430$\pm$140~$\mo$~km~s$^{-1}$,
which is comparable to the linear momentum found in solitary runaway neutron
stars (Mirabel et al.~\cite{mir04}). Other microquasars such as Scorpius X-1
and GRO~J1655$-$40, or the microquasar candidate XTE~J1118+480 (Mirabel
et~al.~\cite{mir01}), display also significant space velocities, although so
far they are not high-energy $\gamma$-ray source candidates. In general, as
suggested by Romero et al.~(\cite{rom04}), some microquasars with high spatial
velocity could be related to the faint, soft and variable unidentified EGRET
sources above and below the Galactic Plane.

\subsection{VHE $\gamma$-ray sources}

The very high energy sky map contains a reduced number of sources. The amount
of confirmed and probable catalogued sources is presently fourteen (6 AGN, 3
pulsar wind nebulae, 3 supernova remnants, 1 starburst galaxy, and 1 unknown)
(Ong~\cite{ong03}). Some microquasars have been observed in the energy range
of TeV $\gamma$-rays with the imaging atmospheric Cherenkov telescopes, but
none of them has been detected with high confidence up to now. Historically,
Cygnus~X-3 was widely observed with the first generation of TeV instruments. 
Some groups claimed that they had detected Cygnus~X-3 (Chadwick
et~al.~\cite{cha85}) whereas other groups failed to detect it (O'Flaherty 
et~al.~\cite{ofl92}). As the claimed detections have not been confirmed, and
the instrumentation at this epoch was limited, these results have not been
considered as definite positive detections by the community. The HEGRA
experiment detected a flux of the order of 0.25~Crab from GRS~1915+105 during
the period May-July 1996 when the source was in an active state (Aharonian \&
Heinzelmann~\cite{aha97}). This source has also been observed with Whipple,
obtaining a 3.1$\sigma$ significance (Rovero et~al.~\cite{rov02}). More
recently, an upper-limit of 0.35 Crab above 400~GeV has been quoted for 
GRS~1915+105 (Horan \& Weekes~\cite{hor03}). LS~I~+61~303 was observed too,
and not detected in the TeV energy range (see Section~\ref{sect:lsi}).





\section{Summary}
\label{sect:discussion}

Microquasars are among the most interesting sources in the Galaxy from the 
viewpoint of high-energy astrophysics. Models predict that radio jets could be
natural sites for the production of high-energy photons via both Compton
scattering and maybe direct synchrotron emission (although hadronic emission
is not discarded either). Although most microquasars have been detected as
soft $\gamma$-ray sources, up to now only a few of them seem to be high-energy
$\gamma$-ray emitters. Future missions such like AGILE and GLAST will confirm
or reject the proposed association between some microquasars and EGRET
sources. Also, observations of microquasars with the new generation of
Cherenkov telescopes (MAGIC, H.E.S.S.) will bring more constraints to the
physics of these systems.

\begin{acknowledgements}
I am grateful to Marc Rib\'o, Valent\'{\i} Bosch-Ramon and Josep Mart\'{\i} for their useful comments and suggestions while writing this paper.
I also acknowledge partial support by DGI of the Ministerio de Ciencia y
Tecnolog\'{\i}a (Spain) under grant AYA2001-3092, as well as partial support
by the European Regional Development Fund (ERDF/FEDER).
\end{acknowledgements}

\label{lastpage}

\end{document}